\documentclass[hyperref,amsmath,amssymb,showpacs,floatfix,journal=jacsat,manuscript=article]{achemso} 
\usepackage{color}
\usepackage{cancel}
\usepackage{graphicx}   % Including figure files
\usepackage{amsmath}    % Advanced maths commands
\usepackage{amssymb}    % Extra maths symbols
\usepackage{extarrows}
\usepackage{morefloats}
\usepackage[version=4]{mhchem} % Formula subscripts using \ce{}
\usepackage{epstopdf}
\usepackage{color}
\usepackage{cancel}
\usepackage{graphicx}   % Including figure files
\usepackage{amsmath}    % Advanced maths commands
\usepackage{amssymb}    % Extra maths symbols
\usepackage{extarrows}
\usepackage{morefloats}
\usepackage[version=4]{mhchem} % Formula subscripts using \ce{}
\usepackage{epstopdf}
\usepackage{mathptmx}
\usepackage{array}
\usepackage{hyperref}
\usepackage{threeparttable}
\usepackage{extarrows}
\DeclareGraphicsExtensions{.jpg, .pdf, .png}
\graphicspath{{./data/}{.}}

\newcommand{\short}{Letter}

\newcommand{\figname}{Figure~}

\newcommand{\figsname}{Figures~}

\newcommand{\gl}{eq}

\newcommand{\Gl}{Eqs}

\newcommand{\alt}{\raisebox{-0.3ex}{$\stackrel{<}{\sim}$}}

\newcommand{\jv}{$J$\textendash$V$}

\SectionNumbersOff
\mciteErrorOnUnknownfalse
\clearpage
\author{Ioan B\^aldea}
\affiliation{Theoretical Chemistry, Heidelberg University, Im Neuenheimer Feld 229, D-69120 Heidelberg, Germany}
\email{ioan.baldea@pci.uni-heidelberg.de}

\title{Dichotomy between Level Broadening and Level Coupling to Electrodes in Large Area EGaIn Molecular Junctions}

\begin{document}

\begin{abstract}
Choosing self-assembled monolayers (SAM) of fluorine terminated oligophenylenes adsorbed on gold as illustration,
    we show that a single level (molecular orbital, MO) model can excellently reproduce full $I$--$V$ curves measured
    for large area junctions fabricated with top EGaIn contact.
    In addition, this model unravels 
    a surprising dichotomy between MO coupling to electrodes and the MO broadening. 
    Importantly for the coherence of the microscopic description, the latter is found to correlate with the
    SAM coverage and molecular and $\pi^\ast$ orbital tilt angles.
\end{abstract}

Interrogating the role of the platform of fabricating molecular junctions
on the transport properties represents an important issue for molecular electronics.
One may ask, for example, whether large area molecular junctions can also be quantitatively described
within the dominant transport channel scenario,\cite{Hush:00,Metzger:01b,Vuillaume:12a,Vuillaume:12c,Baldea:2012g,Fracasso:13,Tao:13,Hou:13,Yu:15,Ho:15,Yutaka:15,Chiechi:15a,Guo:16a,Guo:16b,Lee:16,Lee:16a,Yu:16b,Yu:16d,Lenfant:17,Yu:18,Song:18a,Baldea:2019d,Baldea:2019h,Frisbie:21a,Gu:21,Chiechi:21,Song:22a,Chiechi:22,Jang:23}
which succeeded to provide a coherent microscopical description
of single-molecule junctions obtained via mechanically controlled (MC-BJ) \cite{Zotti:10} and scanning tunneling microscope
(STM-BJ) break junctions\cite{Baldea:2012g} or conducting probe atomic force microscopy
(CP-AFM).\cite{Baldea:2015d,Baldea:2019d,Baldea:2019h,Frisbie:21a}

To analyze transport data for large area molecular junctions, which is the present aim of this {\short},
we will combine results deduced recently \cite{Baldea:2022j,Baldea:2024a}
expressing the tunneling current density $J$ mediated by a single (dominant)
molecular orbital (MO) at different levels of theory yielding the following formulas
\begin{subequations}
\label{eq-J}
  \begin{eqnarray}
  \label{eq-Jex}
  J_{exact} & = & \frac{2 \pi k_B T \sigma}{e \mbox{Re}\,\psi^\prime \left(\frac{1}{2} + i \frac{\varepsilon_0}{2\pi k_B T}\right)}
    \left[
    \mbox{Im}\, \psi\left( \frac{1}{2} + \frac{\Lambda}{2 \pi k_B T} + i \frac{\varepsilon_{V} + e V/2}{2 \pi k_B T}\right) \right . \nonumber \\
    & - & \left .
    \mbox{Im}\, \psi\left( \frac{1}{2} + \frac{\Lambda}{2 \pi k_B T} + i \frac{\varepsilon_{V} - e V/2}{2 \pi k_B T}\right)
    \right]
\end{eqnarray}
\begin{equation}
  J_{\mbox{\small 0K}} = \sigma \frac{\varepsilon_0^2 + \Lambda^2}{e \Lambda}
  \left(\arctan \frac{\varepsilon_{V} + e V/2}{\Lambda} -
  \arctan \frac{\varepsilon_{V} - e V/2}{\Lambda} \right)
  \label{eq-J-0K}
\end{equation}
\begin{equation}
  \label{eq-J-0K-off}
  J_{\mbox{\small 0K,off}} = \sigma\frac{\varepsilon_0^2}{\varepsilon_{V}^2 - (e V/2)^2} V
\end{equation}
\end{subequations}
\begin{equation*}
  \varepsilon_V = \varepsilon_0 + \gamma e V
\end{equation*}
Above, $\psi$ and $\psi^\prime$ are Euler's digamma and trigamma functions of complex argument,\cite{AbramowitzStegun:64}
$\varepsilon_0 $ represents the MO energy offset (which is negative in the specific case of the phenylene-based
molecular junctions where conduction proceeds via the HOMO,\cite{Baldea:2015d} $ \varepsilon_0 = - \vert \varepsilon_0\vert$),
$\gamma$ is the strength of the MO bias driven shift, $\Lambda$ is the (HO)MO width (see below), and 
$T(=298.12$\,K) and $k_B$ are the (room) temperature and Boltzmann's constant, respectively.

The expressions of the low bias conductivity $\sigma$ utilized at arriving at \gl~(\ref{eq-Jex}), (\ref{eq-J-0K}), and (\ref{eq-J-0K-off})
read, respectively\cite{Baldea:2022j}

\begin{subequations}
\begin{eqnarray}
  \label{eq-Gex}
  \sigma_{exact} & = & G_0\frac{\overline{\Gamma}^2}{2\pi\Lambda k_B T}
  \mbox{Re}\,\psi^\prime\left(
  \frac{1}{2} + \frac{\Lambda + i\varepsilon_0}{2\pi k_B T} \right)
\end{eqnarray}
\begin{equation}
  \sigma_{\mbox{\small 0K}} =  G_0 \frac{\overline{\Gamma}^2}{\varepsilon_{0}^2 + \Lambda^2}
  \label{eq-G-0K}
\end{equation}
\begin{equation}
  \label{eq-G-0K-off}
  \sigma_{\mbox{\small 0K,off}} = G_0 \frac{\overline{\Gamma}^2}{\varepsilon_{0}^2}
\end{equation}
\end{subequations}

The quantity
\begin{equation}
  \label{eq-Gamma-bar}
  \overline{\Gamma}^2 = \frac{N_{\mbox{\small eff}}}{\mathcal{A}} \Gamma^2
\end{equation}
is related to the (geometric) average MO coupling $\Gamma = \sqrt{\Gamma_s \Gamma_t}$ to the two
(substrate $s$ and top $t$) electrodes by a factor accounting for the fact that 
in large area molecular junctions with eutectic gallium indium alloy (EGaIn) top electrodes
only a tiny amount of the total number of molecules $N_{\mbox{\small eff}}$
per geometric area $\mathcal{A}$ are current carrying.\cite{Whitesides:14,Frisbie:16d}

As emphasized recently \cite{Baldea:2022j,Baldea:2024a} and exemplary confirmed below,
along with the MO couplings to the two (substrate $s$ and top $t$) electrodes $\Gamma_{s,t}$,
the MO width $\Lambda$ can and does also contain a significant ``extrinsic'' contribution $\Gamma_{env}$
not related with molecule-electrode interactions
\begin{equation}
  \label{eq-Lambda}
  \Lambda = \frac{1}{2}\left(\Gamma_s + \Gamma_t + \Gamma_{env}\right)
\end{equation}

Within the assumption of a single dominant level,
\gl~(\ref{eq-Jex}) is exact for arbitrary biases and temperatures.\cite{Baldea:2024a}
\Gl~(\ref{eq-J-0K}) represents an accurate approximation of \gl~(\ref{eq-Jex}) at sufficiently low temperatures.
\Gl~(\ref{eq-J-0K-off}) accurately reproduces \gl~(\ref{eq-J-0K}) at biases sufficiently far away from resonant tunneling.

Most applications of the single level model to date comprise molecular junctions anchored with thiols
or possessing a hydrogen atom at the physical contact.\cite{Baldea:2012g,Baldea:2015d,Baldea:2019h,Frisbie:21a}
To emphasize the generality of the description based on the single level model,
we will consider as a specific application large area EGaIn-based molecular tunnel junctions
fabricated with the oligophenylene thiol homologous series ($n=1,2,3$) wherein the terminal hydrogen
atom is substituted by a fluorine atom (F-nPT).\cite{Zharnikov:22}

The fact that, out of the various tail group (H$\to$R\,=\,F, \ce{CH3},\,\ce{CF3}) substitutions
investigated experimentally,\cite{Zharnikov:22}
fluorination (\ce{R=F}) yielded the most pronounced changes in the transport properties as compared to the nonsubstituted
oligophenyl thiols (H-nPT) is one reason for this choice. Another reason is the intrinsic electronegativity of the F atom,
which could act as a charge trap at the physisorbed contact,\cite{Kong:15,Zharnikov:22}
possibly giving rise to additional Coulomb interactions at contacts escaping the framework
based on the single level model (referred to as the  ``(non)interacting'' single level model
in the many-body community\cite{AndreiLimits:06,Mehta:06,Doyon:07,Baldea:2012b}).

\begin{figure*}[htb]
  \centerline{
    \includegraphics[width=0.3\textwidth]{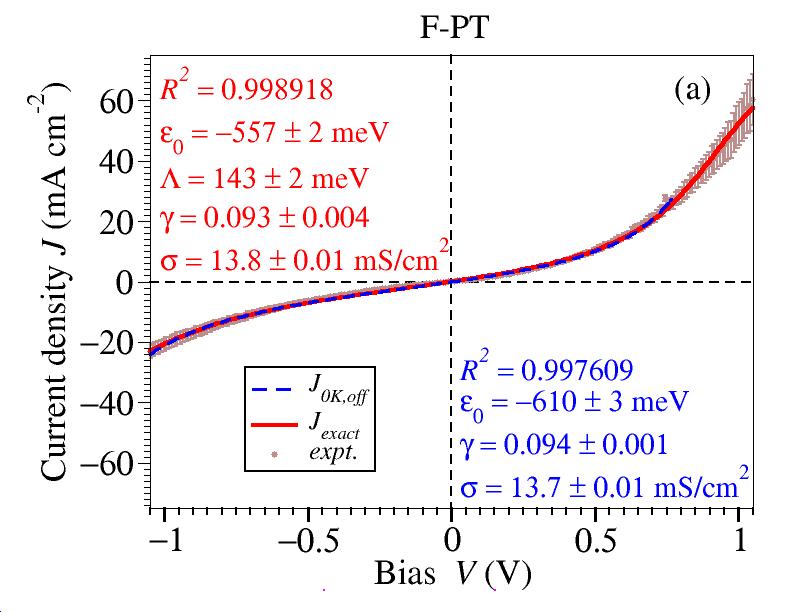}
    \includegraphics[width=0.3\textwidth]{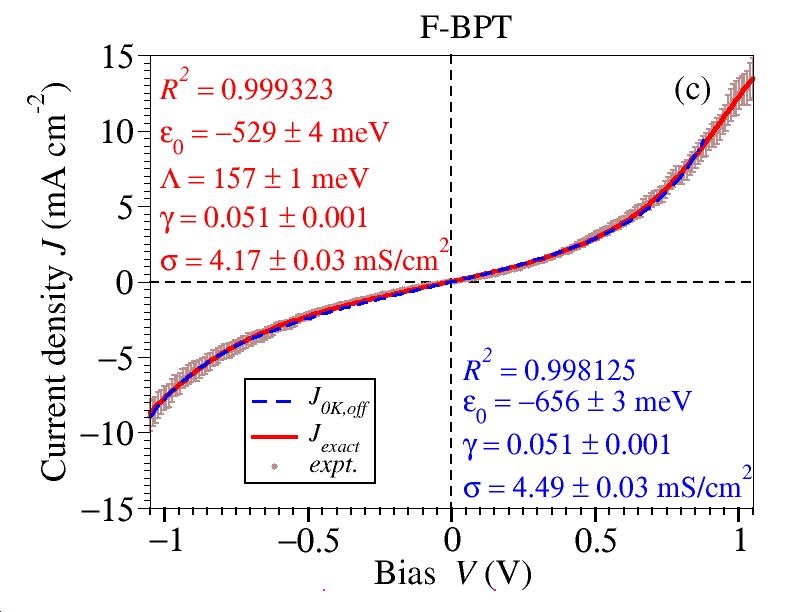}
    \includegraphics[width=0.3\textwidth]{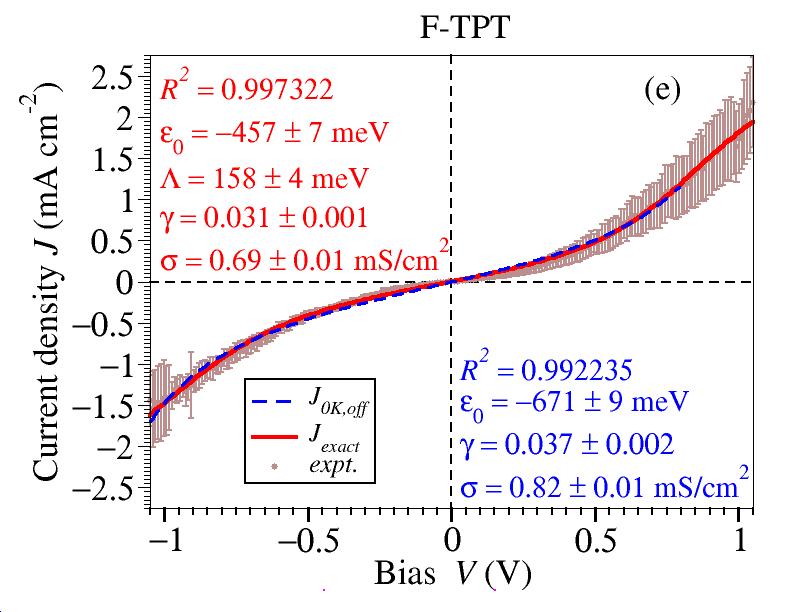}}
  \centerline{
    \includegraphics[width=0.3\textwidth]{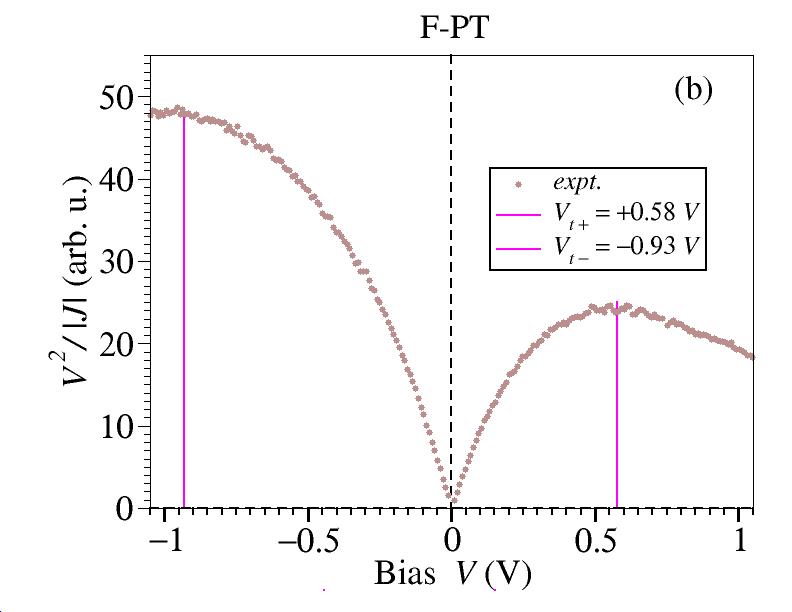}
    \includegraphics[width=0.3\textwidth]{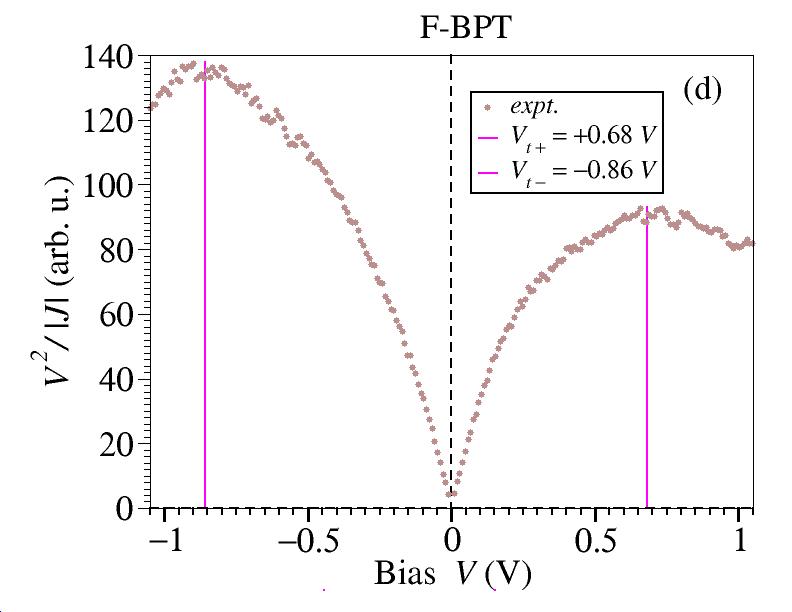}
    \includegraphics[width=0.3\textwidth]{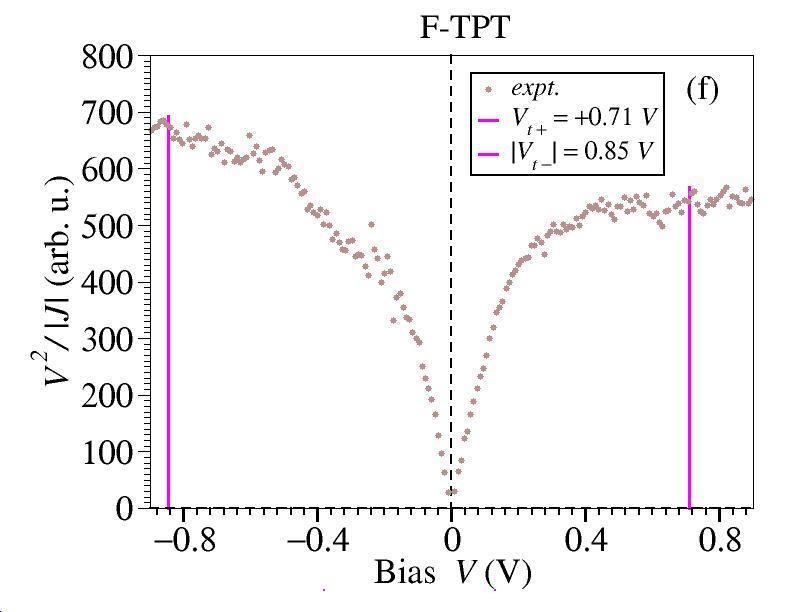}}
  \caption{(a, c, e) Experimental {\jv} curves for
    fluorine terminated large area EGaIn/F-PT-Au, EGaIn/F-BPT-Au, and EGaIn/F-TPT-Au junctions
    containing one, two, or three phenyl rings, respectively
    (brown points, courtesy of Michael Zharnikov) from ref.~\citenum{Zharnikov:22}
    fitted to the exact \gl~(\ref{eq-Jex}) (red lines).
    Fitting curves obtained via \gl~\ref{eq-J-0K}) are not shown because they cannot be distinguished from
    the exact (red) curves within the drawing accuracy.
    Data fitting to \gl~(\ref{eq-J-0K-off}) (results depicted in blue) should employ a bias range
    compatible to this equation ($-1.2\,\vert V_{t-}\vert \alt V \alt 1.25 V_{t+}$), 
    where the transition voltages $V_{t +}$ and $\vert V_{t-}\vert$ for positive and negative biases
    correspond to the maxima of the curve of $V^2/\vert J\vert$ versus $V$ (panels b, d, and f).}
  \label{fig:jv}
\end{figure*}

The parameter values adjusted by fitting the experimental {\jv} data for F-nPT junctions \cite{Zharnikov:22}
to \gl~(\ref{eq-Jex}), (\ref{eq-J-0K}), and (\ref{eq-J-0K-off}) are collected in Table~\ref{table:fit-F-nPT}.
The red curves in \figname\ref{fig:jv} depict the theoretical curves obtained by fitting the experimental {\jv} data (brown points) \cite{Zharnikov:22}
to \gl~(\ref{eq-Jex}) at room temperature ($T = 298.15$\,K).
As visible there, the agreement between the theoretical \gl~(\ref{eq-Jex}) and experiment is excellent.

The fitting curves obtained via \gl~(\ref{eq-J-0K-off}) (assuming $T = 0$\,K) are not shown;
the difference from the exact (red) curves would be invisible within the drawing accuracy of \figname\ref{fig:jv}.
Inspection of Table~\ref{table:fit-F-nPT} reveals that the parameter values based on \gl~(\ref{eq-J-0K} are very
close to those based on \gl~(\ref{eq-Jex}). This demonstrates that, definitely,
thermal effects are altogether negligible for the presently envisaged EGaIn/F-nPT--Au large area junctions.

The fitting curves obtained by using \gl~(\ref{eq-J-0K-off})
are drawn as blue lines in \figname\ref{fig:jv}. We will come back to these curves at a later point in our analysis.

Table~\ref{table:fit-F-nPT} shows that the magnitude of the HOMO energy offset $\vert \varepsilon_0\vert $
slightly decreases with increasing molecular size (number of phenyl rings $n$). This behavior is similar to that reported for
CP-AFM junctions fabricated with (nonsubstituted) oligophenyls,\cite{Baldea:2019d}
but this does not represent the main concern of the present work.
Rather, it is the dependence on the molecular size (number of phenyl rings $n$) of the other model parameters
(namely $\overline{\Gamma}$ and $\Lambda$) that plays a central role in our analysis. 
\begin{table*}
\centering
\caption{\ Model parameter values obtained by fitting the experimental Au-F-nPT/EGaIn  data \cite{Zharnikov:22} to the formula for the current indicated in the first column}
\label{table:fit-F-nPT} 
\begin{tabular*}{0.85\textwidth}{@{\extracolsep{\fill}}rrccc}
  \hline
  Method                 &  Property                  &   F-PT     & F-BPT     & F-TPT  \\
  \hline
  \Gl~(\ref{eq-Jex})     & \(\varepsilon_{0}\)   &  $ -557 \pm 2 $   & $ -529 \pm 4 $ & $ -457 \pm 7 $  \\
                         & \(\Lambda\)          &  $ 143 \pm 2 $   &  $ 157 \pm 1 $ & $ 158 \pm 4 $ \\
                         & \(\gamma\)            &  $ 0.093 \pm 0.004 $ & $ 0.051 \pm 0.001 $ & $ 0.031 \pm 0.001 $ \\
                         & \( \sigma \) (\(\beta = 1.5 \pm 0.2\))  &  $ 13.8 \pm 0.1 $ & $ 4.17 \pm 0.03 $ & $ 0.69 \pm 0.01 $ \\ 
  & \(R^2 \)              & $ 0.998918 $ & $ 0.999323 $ & $ 0.997322 $ \\
    \hline
  \Gl~(\ref{eq-J-0K})    & \(\varepsilon_{0}\)   &  $ -550 \pm 4 $   & $ -519 \pm 3 $ & $ -455 \pm 6 $  \\
                         & \(\Lambda\)          &  $ 149 \pm 2 $   &  $ 163 \pm 1 $ & $ 158 \pm 3 $ \\
                         & \(\gamma\)            &  $ 0.093 \pm 0.001 $ & $ 0.051 \pm 0.001 $ & $ 0.031 \pm 0.001 $ \\
                         & \( \sigma \) (\(\beta = 1.5 \pm 0.2\))  &  $ 14.1 \pm 0.1 $ & $ 4.20 \pm 0.03 $ & $ 0.71 \pm 0.01 $ \\ 
                         & \(R^2 \)              & $ 0.998747 $ & $ 0.99929 $ & $ 0.9971 $ \\
  \hline
  \Gl~(\ref{eq-J-0K-off})  &  \(\varepsilon_{0}\)  & $ -610 $     & $ -656 $ & $ -671 $ \\
                         & \(\gamma\)            &  $ 0.094 $ & $ 0.051 $ & $ 0.037 $ \\
                         & \( \sigma \)  (\(\beta = 1.4 \pm 0.2\))  &  $ 13.7 \pm 0.1 $ & $ 4.49 \pm 0.03 $ & $ 0.82 \pm 0.01 $ \\ 
                         & \(R^2 \)              & $ 0.997609 $ & $ 0.998125 $ & $ 0.992235 $ \\
\hline
Ref.~\citenum{Zharnikov:22} & \(\Sigma\)         & $ 3.9 $ & $ 4.1 $ & $ 4.6 $ \\
                            & \(\theta\)         & $ 48^\circ $ & $ 25.5^\circ $ & $ 21^\circ $ \\
                            & \(\phi\)           & $ 51^\circ $ & $ 68.5^\circ $ & $ 72.3^\circ $ \\
\hline
\end{tabular*}
\end{table*}

\begin{figure*}[htb]
  \centerline{
    \includegraphics[width=0.4\textwidth]{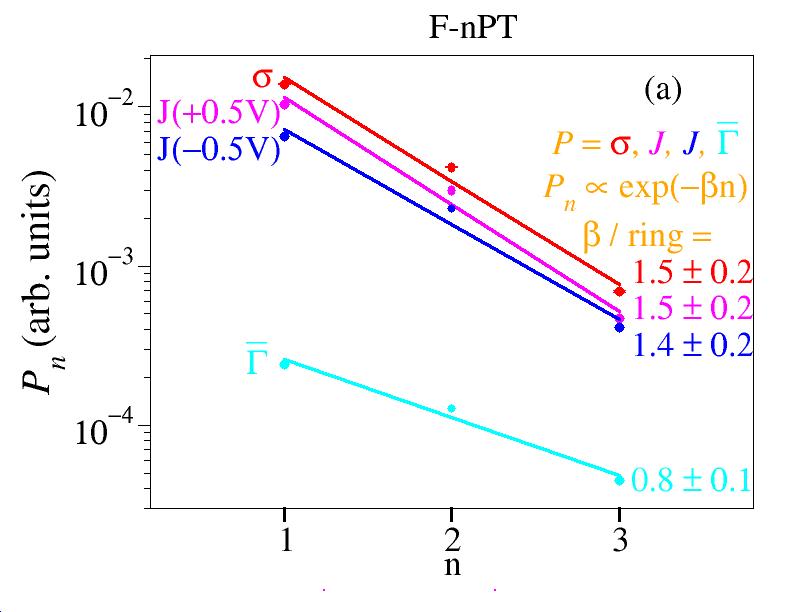}
    \includegraphics[width=0.4\textwidth]{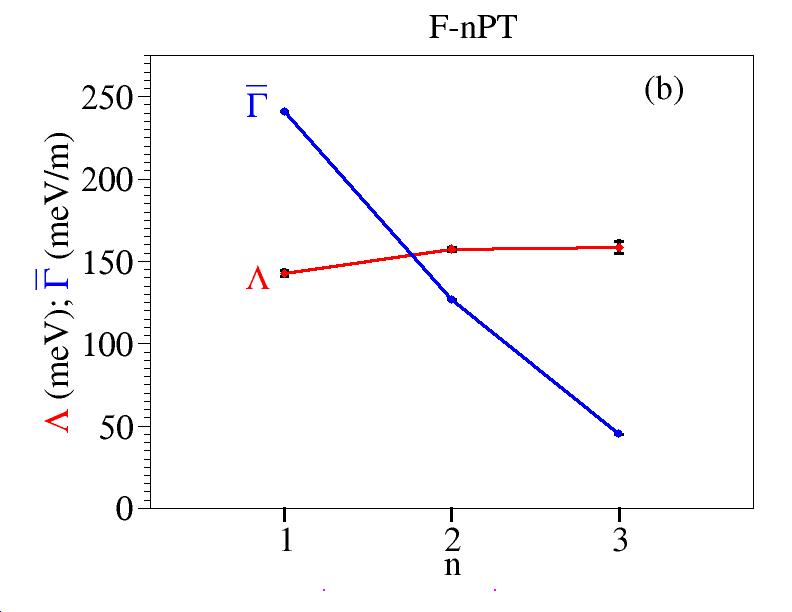}}
  \centerline{ 
    \includegraphics[width=0.4\textwidth]{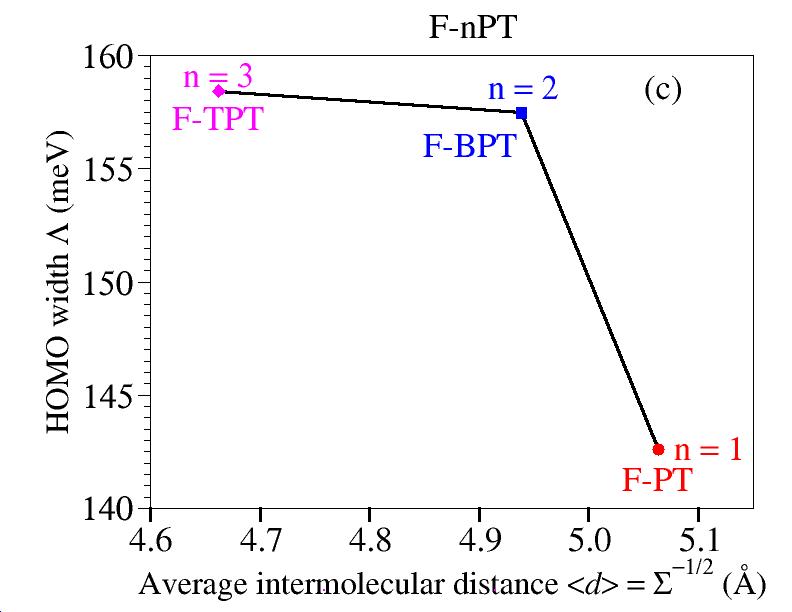}
    \includegraphics[width=0.4\textwidth]{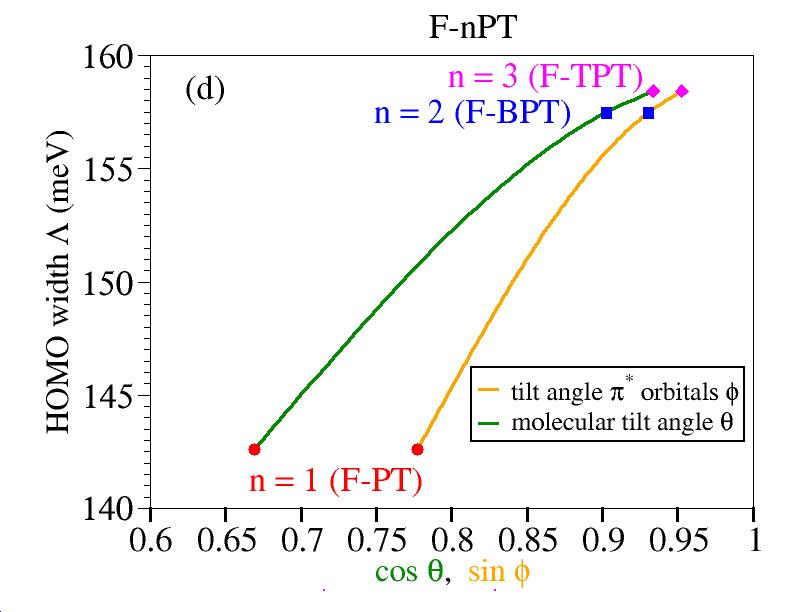}
  }      
  \caption{
    (a) Exponential fall-off with the number of the phenyl rings $n$ of the properties ($P_n$) indicated in the legend: conductivity ($\sigma $),
    current densities $J$ at $\pm 0.5$\,V, and the quantity $\overline{\Gamma} $ proportional to the effective MO coupling to electrodes $\Gamma$.
    (b) The weak dependence of the model parameters $\varepsilon_0 $ and $\Lambda $ on the repeat phenyl units $n$
    contrasting with the strong dependence of $\overline{\Gamma} $ on $n$.
    Dependence of the HOMO width $\Lambda $ (c) on the average intermolecular distance $d$
    expressed in terms of the SAM coverage $\Sigma $
    and (d) on the $\pi^\ast$ orbital tilt angle $\phi$ and molecular tilt angle $\theta$. The lines in panel a represent
    linear data fitting, those of panels b, c and d are guide to the eye.
  }
  \label{fig:Lambda}
\end{figure*}

The marked contrast between $\overline{\Gamma}$ and $\Lambda$ shown in \figname\ref{fig:Lambda}
is an important finding reported in this work.
As visible in \figname\ref{fig:Lambda}a, $\overline{\Gamma}$ exponentially decreases with $n$. Basically,
it is this exponential decay of $\overline{\Gamma}$ that induces the exponential behavior
of the low bias conductivity $\sigma$ (\figname\ref{fig:Lambda}a).
The same applies to the dependence on $n$ of the current densities $J$
at the biases $V = + 0.5$\,V and $V=-0.5$\,V also presented in \figname\ref{fig:Lambda}a,
similar to that already shown in ref.~\citenum{Zharnikov:22}. 

Were the HOMO width $\Lambda $ merely the result of the HOMO-electrode couplings $\Gamma_{s,t}$
(i.e.,~$\Gamma_{env} \equiv 0$ in \gl~(\ref{eq-Lambda})), $\Lambda $ would exhibit a similar exponential decay with $n$
as $\overline{\Gamma}$, which is proportional to $\Gamma_s$ and $\Gamma_t$ (cf.~\gl~(\ref{eq-Gamma-bar})).
However, \figname\ref{fig:Lambda}b reveals that $\Lambda $ is nearly independent of $n$ (for not saying that it even
slightly increases with $n$). This is a clear indication that $\Gamma_s$ and $\Gamma_t$ negligibly contribute to $\Lambda $,
which appears to be dominated by the extrinsic term $\Gamma_{env}$
\begin{equation}
  \label{eq-env-effect}
  \Lambda \approx \Gamma_{env}/2
\end{equation}

Putting in more physical-chemical terms we arrive at another important finding presently reported.
\Gl~(\ref{eq-env-effect}) implies that it is not the HOMO interaction with the electrodes that
that determines the HOMO broadening but rather the interaction of a given molecule with its neighboring molecules
of the SAM.

If so, $\Lambda $ should be correlated with SAM properties. To demonstrate that this is indeed the case,
we plotted in \figname\ref{fig:Lambda}c the values of $\Lambda$ deduced by fitting the measured {\jv} curves (Table~\ref{table:fit-F-nPT})
against the average intermolecular distance $\langle d \rangle = \Sigma^{-1/2}$ expressed using the experimental values of the SAM
coverage $\Sigma$ (\figname\ref{fig:Lambda}c).
We also show plots of $\Lambda$ against the average tilt angle of the
$\pi^\ast$ orbitals $\phi$ and the average molecular tilt angle $\theta$ (both expressed with respect to the
normal at surface) (\figname\ref{fig:Lambda}d).

The monotonic dependencies
of \figsname\ref{fig:Lambda}c and \ref{fig:Lambda}d convey an important message: more distant molecules in the SAM weaken
the intra-SAM interactions which become less effective in broadening the HOMO density of states.
This result
provides microscopic support for the fact that, in the molecular junctions considered,
the HOMO broadening is not dominated by the molecule-electrode interactions but rather by the
intra-SAM interactions between neighboring molecules.

Before ending, a comment on the fitting curves obtained by using \gl~(\ref{eq-J-0K-off})
drawn as blue lines in \figname\ref{fig:jv} is in order. 
As seen in \figsname\ref{fig:jv}a, c, and e, these (blue) curves also reproduce well the experimental measurements
in the bias range ($ -1.2 \vert V_{t -}\vert \alt V \alt  1.2\,V_{t +}$) where \gl~(\ref{eq-J-0K-off}) applies.\cite{Baldea:2012a,Baldea:2023a}
Here, $V_{t +}$ and $ V_{t -}$ are transition voltage for positive and negative bias polarities,
which are defined by the maxima of the curves for $V^2/\vert J\vert $ (\figsname\ref{fig:jv}b, d, and f).
As visible in Table~\ref{table:fit-F-nPT} and \figsname\ref{fig:jv}a, c, and e,
the MO offset $\vert \varepsilon_0\vert $ obtained using \gl~(\ref{eq-J-0K-off})
are somewhat overestimated compared to those obtained using the exact \gl~(\ref{eq-Jex}).
The reason is that,
to be accurate, the off-resonance condition underlying \gl~(\ref{eq-J-0K-off}) does not only imply
biases sufficiently away from resonance ($ e \vert V\vert = 2 \vert \varepsilon_V\vert $) but also
$\Lambda \alt \vert \varepsilon_0 \vert / 10 $.\cite{Baldea:2023a}
The latter condition is not satisfied by the values of Table~\ref{table:fit-F-nPT},
and for this reason, the estimates for $\varepsilon_0$ based \gl~(\ref{eq-J-0K-off})
are not so accurate as in many other cases investigated earlier.
\cite{Vuillaume:12a,Vuillaume:12c,Fracasso:13,Tao:13,Hou:13,Yu:15,Ho:15,Yutaka:15,Chiechi:15a,Guo:16a,Guo:16b,Lee:16,Lee:16a,Yu:16b,Yu:16d,Lenfant:17,Yu:18,Song:18a,Baldea:2019d,Baldea:2019h,Frisbie:21a,Gu:21,Chiechi:21,Song:22a,Chiechi:22,Jang:23}

In closing, in this {\short} we demonstrated
that, notwithstanding the potential action as charge trap of the terminal F atom
they contain,\cite{AndreiLimits:06,Mehta:06,Doyon:07,Baldea:2012b} large area
molecular junctions fabricated with fluorine terminated oligophenyls and EGaIn top electrodes
(i) exhibit transport properties that can be excellently reproduced
using a recently deduced analytic formula for the current based on a single level (MO) model,
which reveals (ii) that HOMO-electrode coupling $\Gamma$
and the HOMO broadening $\Lambda$ have different physical-chemical origins,
and (iii) that, rather than being dominated by molecule-electrode interaction,
the HOMO broadening is basically due to intra-SAM interactions,
as witnessed by the correlation of $\Lambda$ with the properties
of the SAM (coverage and tilt angles).

From a methodological perspective, the present work is a plea in favor of investigating junctions fabricated 
using homologous molecular series rather than disparate molecular species.
In arriving at conclusions (ii) and (iii) mentioned above, pursuing this route
was essential.

This research did not receive any specific financial support
but benefited from computational support by the
state of Baden-W\"urttemberg through bwHPC and the German Research Foundation through
Grant No.~INST 40/575-1 FUGG (bwUniCluster 2, bwForCluster/HELIX, and JUSTUS 2 cluster).

\section*{Conflicts of interest}
There are no conflicts to declare.
\renewcommand\refname{References}
\providecommand{\latin}[1]{#1}
\makeatletter
\providecommand{\doi}
  {\begingroup\let\do\@makeother\dospecials
  \catcode`\{=1 \catcode`\}=2 \doi@aux}
\providecommand{\doi@aux}[1]{\endgroup\texttt{#1}}
\makeatother
\providecommand*\mcitethebibliography{\thebibliography}
\csname @ifundefined\endcsname{endmcitethebibliography}
  {\let\endmcitethebibliography\endthebibliography}{}

\end{document}